\documentstyle[prb,aps,epsfig]{revtex}

\begin{document}
\draft
\title{Quantitative Imaging of Sheet Resistance with a Scanning Near-Field
Microwave Microscope}
\author{D. E. Steinhauer,$^{a)}$\footnotetext{$^{a)}$
Electronic mail: steinhau@squid.umd.edu. Color versions of the figures in
this paper can be found at http://www.csr.umd.edu/research/hifreq/micr\_microscopy.html.} C. P. Vlahacos, S. K. Dutta, B. J. Feenstra, F. C. Wellstood, and Steven M. Anlage}
\address{Center for Superconductivity Research, Department of Physics, University of\\
Maryland, College Park, MD 20742-4111 }
\maketitle

\begin{abstract}
We describe quantitative imaging of the sheet resistance of metallic
thin films by monitoring frequency shift and quality factor in a
resonant scanning near-field microwave microscope. This technique allows
fast acquisition of images at approximately 10 ms per pixel over a frequency
range from 0.1 to 50 GHz. In its current configuration, the system
can resolve changes in sheet resistance as small as 0.6 $\Omega /\Box $ for
100 $\Omega /\Box $ films. We demonstrate its use at 7.5 GHz by generating a
quantitative sheet resistance image of a YBa$_2$Cu$_3$O$_{7-\delta }$ thin
film on a 5 cm-diameter sapphire wafer.
\end{abstract}

\pacs{}

Non-destructive imaging of microwave sheet resistance has been demonstrated
using a variety of probes combined with resonant systems. Far-field techniques,
such as confocal\cite{Hogan,Martens} and dielectric\cite{Gallop,Wilker} resonators,
while allowing quantitative sheet resistance imaging, have the disadvantage
of relatively low spatial resolution ($\gtrsim $1 mm) limited by the
wavelength. Near-field microscopy using coaxial,\cite{Bryant,Takeuchi}
microstrip,\cite{Tabib-Azar} or waveguide\cite{Golosovsky} resonators,
offers higher spatial resolution. However, current techniques either require
contact with the sample,\cite{Takeuchi} inhibiting quantitative
interpretation of the data, have low sheet resistance sensitivity in
conducting samples,\cite{Bryant,Tabib-Azar,Xu} or require additional image
processing to obtain sub-wavelength resolution.\cite{Golosovsky} For
non-destructive sheet resistance imaging of thin films, it is desirable to
have quantitative methods that combine high resolution, high speed, simple
construction from commercially-available components, and straightforward
image interpretation. We describe here the application of an open-ended
coaxial probe resonator to obtain quantitative images of microwave sheet
resistance with $\lambda $/80 spatial resolution.

\begin{figure}[htb]
\begin{center}
\leavevmode
\epsfxsize=8cm
\epsffile{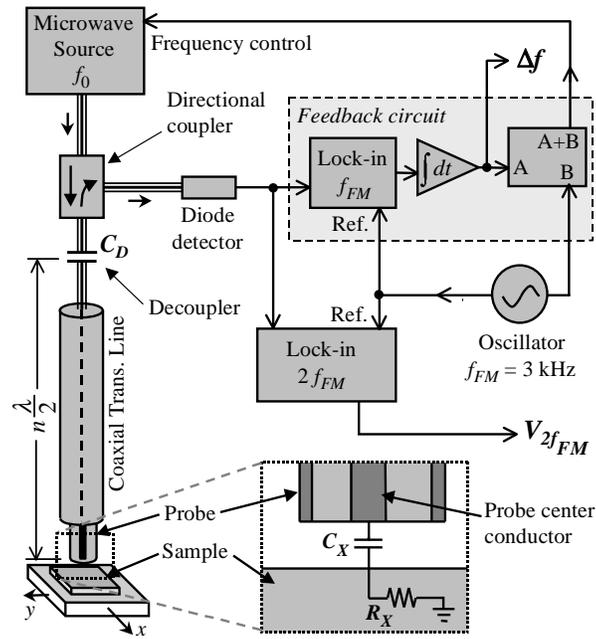}
\end{center}
\caption{Schematic of the scanning near-field microwave microscope. The inset shows the
interaction between the probe and the sample, represented by a capacitance $C_{X}$ and a resistance $R_{X}$.}
\label{schematic}
\end{figure}

Our scanning microwave microscope consists of a resonant coaxial
transmission line connected to an open-ended coaxial probe and a microwave
source (Fig.\ \ref{schematic}).\cite{Vlahacos,Anlage,Steinhauer} The
microwave source, which is weakly coupled to the resonant transmission line
through a decoupling capacitor $C_D$, is frequency modulated by an external
oscillator at a rate $f_{FM}\sim $ 3 kHz. The electric field at the probe
tip is perturbed by the region of the sample beneath the probe's center
conductor. We monitor these perturbations using a diode detector which
produces a voltage signal proportional to the power reflected from the
resonator. A feedback circuit\cite{Steinhauer} (Fig.\ \ref{schematic}) keeps the microwave source
locked to a resonant frequency of the transmission line, and has a voltage
output which is proportional to shifts in the resonant frequency due to the sample.

To determine the quality factor Q of the resonant circuit, a lock-in
amplifier, referenced at $2f_{FM}$, gives an output voltage $V_{2f_{FM}}$
which is related to the curvature of the reflected power-vs.-frequency curve
on resonance, and hence to Q. To relate $V_{2f_{FM}}$ and Q, we perform a
separate experiment, in which we vary Q using a
microwave absorber at various heights below the probe tip, and measure the absolute
reflection coefficient $\left| \rho \right| ^2$ of the resonator. If $\left| \rho _0\right| ^2$ is the reflection
coefficient at a resonant frequency $f_0$, then the coupling coefficient between the source and the resonator is $%
\beta =\left( 1-\left| \rho _0\right| \right) /\left( 1+\left| \rho
_0\right| \right) .$ The loaded quality factor of the resonator\cite{Aitken,Zaki} is Q$_L$ $%
=f_0/\Delta f$, where $\Delta f$ is the difference in frequency between the two points where $\left| \rho
\right| ^2=$ $\left( 1+\beta ^2\right) /\left( 1+\beta \right) ^2$. The unloaded quality factor, which is the Q of the resonator
without coupling to the microwave source and detector, is then Q$_0=$ Q$%
_L\left( 1+\beta \right) $. We also measure $V_{2f_{FM}}$, and find that
there is a unique functional relationship between Q$_0$ and $V_{2f_{FM}}$;
thus, we need to calibrate this relationship only once for a given
microscope resonance. In a typical scan, we record $V_{2f_{FM}}$, and
afterward convert $V_{2f_{FM}}$ to Q$_0$.

\begin{figure}[htb]
\begin{center}
\leavevmode
\epsfxsize=8cm
\epsffile{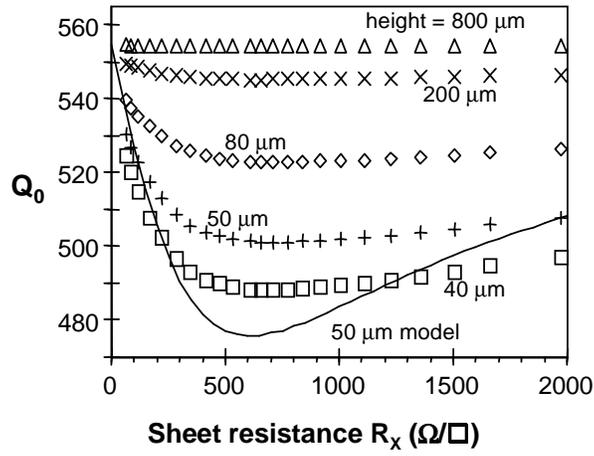}
\end{center}
\caption{Unloaded quality factor Q$_{0}$ of the resonant circuit as a
function of the sheet resistance $R_{X}$ of a variable-thickness aluminum
thin-film sample. The labels indicate different probe-sample separations. A
probe with a 500 $\mu$m center conductor was used at a frequency of 7.5 GHz.
The solid line indicates a model calculation for a probe-sample separation
of 50 $\mu$m. }
\label{Qgraph}
\end{figure}

To determine the relationship between Q$_0$ and sample sheet resistance ($%
R_X $), we used a variable-thickness aluminum thin film on a glass substrate.%
\cite{Steinhauer} The cross-section of the thin film is wedge-shaped,
implying a spatially varying sheet resistance. Using a probe with a 500 $\mu 
$m diameter center conductor, and selecting a resonance of the microscope
with a frequency of 7.5 GHz, we acquired frequency-shift and Q$_0$ data. We then cut the sample into narrow strips to take two-point resistance measurements and determine the
local sheet resistance. The unloaded Q$_0$ of the resonator as a function of 
$R_X$ is shown in Fig.\ \ref{Qgraph} for various probe-sample separations.
We note that Q$_0$ reaches a maximum as $R_X\rightarrow 0$; as $R_X$
increases, Q$_0$ drops due to loss from currents induced in
the sample, reaching a minimum around $R_X=660\ \Omega /\Box $ for a height
of 50 $\mu $m. Similarly, as $R_X\rightarrow \infty $, Q$_0$ increases 
due to diminishing currents in the sample.

We also note that when the probe is located 50 $\mu $m above the bare glass
substrate, Q$_0=549$, which is only slightly less than Q$^{\prime}$ = 555 when the
probe is far away ($>$1 mm) from the sample. Using $1/$Q$_0=1/$Q$_s+1/$Q$^{\prime}$, we find Q$
_s=51000\gg $ Q$^{\prime }=555$, where Q$_s$ is associated with losses in
the glass substrate, and Q$^{\prime}$ is associated with losses in the
transmission line. As a result, we conclude that the glass substrate has
little effect on Q$_0$. In contrast, because the dielectric substrate effectively lengthens the microscope resonant circuit, frequency shift is highly sensitive to the
substrate.\cite{Steinhauer,Vlahacos2} This suggests that we use the Q data, rather than the frequency shift data, to generate substrate-independent images of thin film sheet resistance.

As shown in Fig.\ \ref{Qgraph}, $R_X$ is a double-valued function of Q$_0$.
This presents a problem for converting the measured Q$_0$ to $R_X$. However, 
$R_X$ is a single-valued function of the frequency shift,\cite{Steinhauer}
allowing one to use the frequency shift data to determine which branch of
the $R_X($Q$)$ curve should be used.

To better understand the system's behavior, we modeled the interaction
between the probe and sample as a capacitance $C_X$ between the probe's
center conductor and the sample in series with the sample sheet resistance $R_X$ (see Fig.\ \ref{schematic}).\cite{Steinhauer,Vlahacos2} We assumed a parallel
plate approximation for the capacitor $C_X$, and took $R_X=\tilde{\rho} /t$,\cite
{Steinhauer} where $\tilde{\rho }$ is the sample dc resistivity, and $t$ is the film
thickness. To find Q$_0$, we calculated the width of the resonant minimum in
the reflection coefficient $\left| \rho \right| ^2$ vs. frequency curve,
as described above. The model results for a height of 50 $\mu $m are
indicated by the solid line in Fig.\ \ref{Qgraph}. To fit the data we used
the measured Q$_0=555$ and Q$_L$ $=353$ with the probe far away from the
sample to fix two fitting parameters: the coupling capacitance $C_D=$ 0.17
pF, and the transmission line attenuation constant $\alpha =-$1.26 dB/m (in
close agreement with the manufacturer's specified value\cite{Gore} of $%
\alpha =-$1.23 dB/m). Aside from a coupling factor involving $C_X$, the
model agrees qualitatively and predicts the $R_X$ which yields the minimum Q$%
_0$ to within 10\%. For large $R_X$, the model predicts a faster return to
the asymptotic Q$_0$ value of 555. The overall behavior of this
simple model confirms our understanding of the system.

To explore the capabilities of our system, we scanned a thin film of YBa$_2$Cu$_3$O$_{7-\delta }$
(YBCO) on a 5 cm-diameter sapphire substrate at room temperature. The film
was deposited using pulsed laser deposition with the sample temperature
controlled by radiant heating. The sample was rotated about its center
during deposition, with the $\sim $3 cm diameter plume held at a position
halfway between the center and the edge. The
thickness of the YBCO thin film varied from about 100 nm at the edge to 200 nm near the center.

\begin{figure}[htb]
\begin{center}
\leavevmode
\epsfxsize=8cm
\epsffile{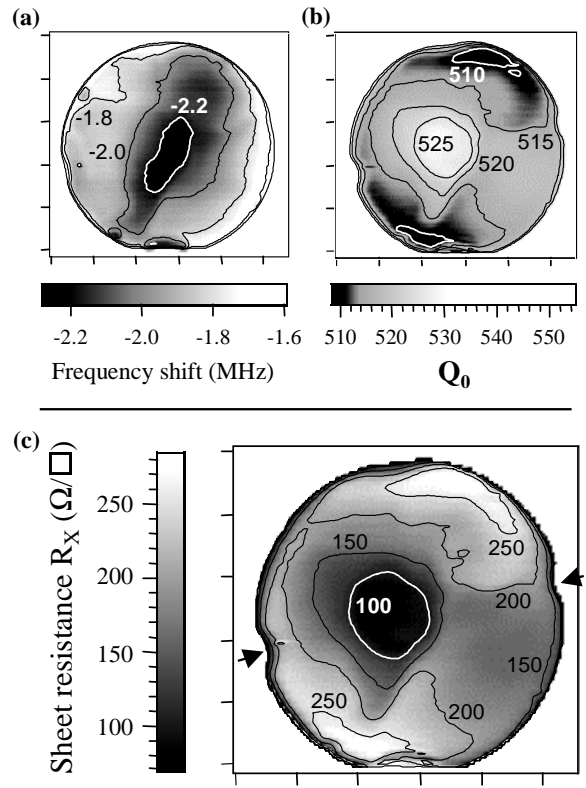}
\end{center}
\caption{Images of a variable-thickness YBCO thin-film on a 5 cm-diameter
sapphire wafer, where the film is the thickest at the center. The tick marks
are 1 cm apart for the images of (a) frequency shift relative to the
resonant frequency when the probe is far away ($>$1 mm) from the sample, (b)
unloaded Q, and (c) sheet resistance ($R_{X}$). The arrows in (c) point to
small semi-circular regions where clips held the wafer during deposition,
and thus no film is present. The labels indicate values at each contour
line. A probe with a 500 $\mu$m diameter center conductor was used at a
height of 50 $\mu$m, at a frequency of 7.5 GHz. }
\label{YBCOimages}
\end{figure}

Figure\ \ref{YBCOimages} shows three microwave images of the YBCO sample.
The frequency shift [Fig.\ \ref{YBCOimages}(a)] and Q$_0$ [Fig.\ \ref{YBCOimages}(b)] were acquired
simultaneously, using a probe with a 500 $\mu $m-diameter center conductor
at a height of 50 $\mu $m above the sample. The scan took approximately 10
minutes to complete, with raster lines 0.5 mm apart. The frequency shifts in
Fig. \ref{YBCOimages}(a) are relative to the resonant frequency of 7.5 GHz
when the probe was far away ($>$1 mm) from the sample; the resonant frequency shifted downward by more than
2.2 MHz when the probe was above the center of the sample. Noting that the
resonant frequency drops monotonically between the edge and the center of
the film, and that the resonant frequency is a monotonically increasing function of sheet resistance,\cite{Steinhauer} we conclude that the sheet resistance decreases monotonically
between the edge and the center.

The frequency shift and Q$_0$ images [Fig.\ \ref{YBCOimages}(a) and (b)]
differ slightly in the shape of the contour lines. This is most likely due
to the 300 $\mu $m-thick substrate being warped, causing a variation of a
few microns in the probe-sample separation during the scan. However, for
a sample such as that shown in Fig.\ \ref{YBCOimages}, with an $R_X$ variation across the sample of $\sim 100\ \Omega /\Box $, the Q$_0$ data are
primarily sensitive to changes in $R_X$, while the frequency shift data are
primarily sensitive to changes in probe-sample separation. As a result, we
attribute the difference between the frequency shift and Q$_0$ images to
small changes in probe-sample separation, which will mainly affect the
frequency shift data. Since the values of $R_X$ are retrieved from the Q
data, the warping does not affect the final $R_X$ appreciably. In principle,
one can use the combined frequency shift and Q images to extract sample
topography information,\cite{Vlahacos2} allowing one to separate the effects
of sample sheet resistance and topography.

From Fig.\ \ref{YBCOimages}(b), we see that the lowest Q occurs near the
edge of the film, and that the Q rises toward the center of the sample. As
mentioned above, $R_X$ is not a single-valued function of Q and we must use the frequency shift image [Fig.\ \ref{YBCOimages}(a)] to determine which
branch of the Q$_0$ vs.\ $R_X$ curve in Fig.\ \ref{Qgraph} to use. From the
frequency shift image we learned that $R_X$ decreases monotonically from the
edge to the center of the sample; therefore we use the branch of the Q$_0$
vs.\ $R_X$ curve with $R_X<660\ \Omega /\Box $, since this is the branch
that yields a decreasing $R_X$ for increasing Q$_0$.

With the appropriate branch identified, we then transformed the Q image in Fig.\ 
\ref{YBCOimages}(b) to the sheet resistance image in Fig.\ \ref{YBCOimages}%
(c) using a polynomial least-squares fit to the data presented in Fig.\ \ref{Qgraph} for R$_X<540\ \Omega /\Box $ and a height of 50 $\mu $m. Figure\ \ref{YBCOimages}(c) confirms that the film does indeed have a
lower resistance near the center, as was intended when the film was
deposited. We note that the sheet resistance does not have a simple radial
dependence, due to either non-stoichiometry or defects in the film.

After scanning the YBCO film, we patterned it and made four-point dc resistance measurements all over the wafer.  The dc sheet resistance had a spatial dependence identical to the microwave data in Fig.\ \ref{YBCOimages}(c). However, the absolute values were approximately twice as large as the microwave results, most likely due to degradation of the film during patterning.

To estimate the sheet resistance sensitivity, we monitored the noise in $%
V_{2f_{FM}}$. We find the Q sensitivity of the system to be $\Delta $Q$_0$ $%
\approx 0.08$ for Q$_0$ = 555 and an averaging time of 10 ms. Combining this
with the data in Fig.\ \ref{Qgraph}, we find $\Delta R_X/R_X=6.4\times
10^{-3}$, for $R_X=100\ \Omega /\Box $ using a probe with a 500 $\mu $m
diameter center conductor at a height of 50 $\mu $m and a frequency of 7.5
GHz. The sensitivity scales with the capacitance between the probe center
conductor and the sample ($C_X$); increasing the diameter of the probe
center conductor and/or decreasing the probe-sample separation would improve
the sensitivity.

In conclusion, we have demonstrated a technique which uses a near-field
microwave microscope to generate quantitative sheet resistance images of
thin film samples. The strengths of our system include the ability to arrive
at quantitative results and to confirm our understanding of the system with
a simple model. Other advantages include its speed, measurement frequency
bandwidth, construction from standard commercially-available components, and
the possibility of enhancing its spatial resolution by using a probe with a
smaller-diameter center conductor.\cite{Vlahacos}

The authors would like to thank Alberto Pique of Neocera, Inc., for the YBCO
wafers. This work has been supported by NSF-MRSEC grant No. DMR-9632521, NSF
grants No. ECS-9632811 and DMR-9624021, and by the Center for
Superconductivity Research.

\newpage

\newpage

\newpage

\end{document}